\documentclass[9pt,twocolumn,twoside]{opticajnl}
\journal{opticajournal} 

\usepackage{appendix}
\setboolean{shortarticle}{true}

\usepackage{lineno}
\usepackage{appendix}

\title{Adaptive optical signal-to-noise ratio recovery for long-distance optical fiber transmission}

\author[1]{Mingwen Zhu}
\author[1]{Shangsu Ding}
\author[1]{Zhixue Li}
\author[1]{Song Yu}
\author[1]{Jianming Shang}
\author[1,*]{Bin Luo}
\affil[1]{State Key Laboratory of Information Photonics and Optical Communications, Beijing University of Posts and Telecommunications, Beijing, 100876, China}

\affil[*]{luobin@bupt.edu.cn}

\begin{abstract}
In long-distance fiber optic transmission, the optic fiber link and erbium-doped fiber amplifiers can introduce excessive noise, which reduces the optical signal-to-noise ratio (OSNR). The narrow-band optical filters can be used to eliminate noise and thereby improve OSNR. However, there is a relative frequency drift between the signal and the narrow-band filter, which leads to filtered signal instability. This paper proposes an adaptive OSNR recovery scheme based on a Fabry-Perot (F-P) cavity with mode width of 6 MHz. Utilizing the comb filtering of F-P cavity, the noise around the carrier and sidebands of the signal is filtered out simultaneously. To avoid frequency mismatch, we propose a double-servo scheme to suppress relative frequency drift between the signal and the F-P cavity. We constructed a stable radio frequency transfer system based on passive phase compensation and compared our scheme with other OSNR recovery schemes based on optical filters. Compared to the schemes based on dense wavelength division multiplexing (DWDM) and Waveshaper, our scheme demonstrates an improvement in OSNR of carrier by at least 12 dB and sidebands by at least 23.5 dB. The short-term transfer stability (1 s) is improved by one order of magnitude compared to DWDM and half an order of magnitude compared to Waveshper. This scheme can be applied to the recovery of signals with low OSNR in long-distance fiber optic transmission, improving signal quaility and extending the transmission distance limit. 
\end{abstract}

\setboolean{displaycopyright}{false} 

\begin{document}

\maketitle

\section{Introduction}

Fiber optic transmission systems, such as time and frequency transfer and optical communications, are widely used in commercial and scientific applications. They are crucial for modern telecommunications \cite{yu2010ultra}, navigation \cite{tseng2013survey}, and geodesy \cite{grotti2018geodesy}. For a long time, the capacity and coverage of optical fiber transmission systems are limited by Kerr nonlinearity, which limits the signal transmit power and ultimately determines the maximum achievable link optical signal-to-noise ratio (OSNR) \cite{essiambre2012capacity}.
\par In long-distance fiber optic transmission systems, the signal intensity gradually decreases with increasing transmission distance. If power compensation is not performed in the middle node of the link, the OSNR of the received signal will still be very low, even if high-gain optical amplification is used at the end. Currently, there are two solutions to this problem: power configuration and noise filtering. In terms of power configuration, optical amplifiers should be placed in the middle node of the link to compensate for signal light attenuation \cite{predehl2012920,droste2013optical}. In fiber optic transmission systems the following types of optical amplifiers are used: semiconductor optical amplifiers (SOAs), Raman amplifiers, Brillouin amplifiers, optical injection-locked (OIL) amplifiers, Brillouin amplifiers, and erbium-doped fiber amplifiers (EDFAs). The main problem with SOAs is that their noise figure is high and their gain dynamics can cause serious signal distortions \cite{bobrovs2013comparative}. The Brillouin amplifier provides higher gain for low power signals, but its gain spectrum width is narrow and the frequency of the modulation signal is limited \cite{Terra:10}. Raman amplifiers offer low noise and large bandwidth, with a low noise figure. The use of Raman amplifiers in fiber optic transmission can effectively reduce the degradation of OSNR \cite{islam2002raman}. However, their amplification gain is low, making it difficult to meet the power compensation demands for long distance transmission \cite{okoumassoun2023performance}. OIL amplification provides high gain and low noise. However, it has a narrow amplification bandwidth and requires a feedback loop for frequency control to maintain the injection-locked state \cite{liu2020optical}. Additionally, it is typically limited to amplifying a single frequency, which may not be suitable for many applications. EDFAs are commonly used due to their simple structure and high gain \cite{pedersen1991design}. However, each EDFA introduces amplified spontaneous emission (ASE) \cite{demir2007nonlinear}. This causes the OSNR to decrease continuously with the transmission distance, resulting in a decreased electrical signal-to-noise ratio (SNR) and increased frequency fluctuation of the received signal \cite{charles1966some}. In certain applications that require high signal stability, it is often necessary to utilize the phase-locked loop to regenerate the signal \cite{liu2020ultrastable, chiou1996effect}. Low SNR signals may result in frequent cycle hopping or even loss of lock \cite{ascheid1982cycle}. Although it is possible to optimize the gain and distribution of amplifiers in the link to improve the SNR, the large number of amplifiers used over long distances makes it time-consuming and laborious to adjust each amplifier individually. \cite{salwik2021optimization,sliwczynski2012bidirectional}. Especially in outdoor transmission, remote control of amplifier gain is also limited by network conditions. Adding optical filters at the receiver can increase the OSNR and filter out noise. This method avoids the need for tedious remote control. Currently, the most common way of long-distance optical transmission is the utilisation of EDFAs as a intermediate node amplifier with optical filtering receiver end \cite{yin2018reference,weiss2003performance}.

\par Optical filters can effectively filter out optical noise from EDFAs. Currently, common optical filters in fiber optic transmission systems are dense wavelength division multiplexing (DWDM) \cite{aloisio2012performance}, Waveshaper \cite{gavioli2010investigation}, fiber bragg grating (FBG) \cite{ozolicnvs2009realization}, and comb filter (e.g., Fabry-Perot cavity) \cite{iodice2000silicon}. However, the filtering bandwidths of DWDM, Waveshaper, and FBG filters cannot be very narrow. Although the filtering bandwidths of Waveshaper can be as low as 10 GHz \cite{gavioli2010investigation}, the laser linewidth are usually at or below the MHz level, resulting in residual noise even after filtering. The filtering bandwidth of the comb filter can be narrow enough, but its center frequency and longitudinal mode spacing are not stable \cite{kim2016rapidly}. This can cause the wavelength of the signal light to easily drift out of the range of the filter combs , resulting in an unstable or even disappearing received signal \cite{Yu:24}.
\par This paper proposes an adaptive OSNR recovery scheme that uses the filtering characteristics of the comb of the Fabry-Perot (F-P) cavity to eliminate noise outside the signal carrier and sidebands. The chosen F-P cavity has a mode width ($\Gamma _{FWHM}$) of only 6 MHz and a high extinction ratio, which can filter out most of the noise. For the signal carrier and sidebands wavelength drift relative to F-P cavity longitudinal mode, we propose a double-loop locking scheme to lock the laser center frequency to the F-P cavity transmission peak and simultaneously stabilize the F-P cavity longitudinal mode spacing so that both carrier and sidebands can pass through the F-P cavity. Our experiments demonstrate that this feedback system has excellent long-term stability. To evaluate filtering performance, we established an RF transfer system based on passive compensation. This system introduces significant attenuation to simulate the extremely low OSNR condition of the signal after passing through a long fiber link. Compared to DWDM (bandwidth 100 GHz) and Waveshaper (bandwidth 10 GHz), the F-P cavity filtering can improve the OSNR of carrier and sideband by approximately 12 dB and 23.5 dB. The OSNR is close to that of the signal after modulation. The ALLAN deviation (ADEV) of the F-P cavity filter exhibiting an order of magnitude and a half-order of magnitude improvement over DWDM and Waveshaper at average time of 1 s, respectively. It demonstrates superior short-term stability. The proposed F-P cavity filtering scheme can be combined with the traditional DWDM system, and the OSNR can be greatly improved by adding it to the middle or end nodes of the link. This scheme can also be combined with other OSNR optimization methods to further improve the quality of the received signal. In the future, it can be integrated on the chip based on this scheme to further improve the compactness and convenience of use.
\section{Principle}
\subsection{OSNR}

In long-distance optical fiber transmission systems, the OSNR and SNR of the signal gradually decrease with the transmission distance, reducing the precision of transmitted information. Assuming a unidirectional fiber transmission system of N EDFAs cascaded, and only one wavelength is transmitted in a fiber link. In this case, we only consider the ASE introduced by EDFA. The input signal optical power is $P_{in}$, and the output signal optical power is $P_{out}$. The total gain of EDFAs in the fiber link is $G$. The total noise power of the ASE after cascading EDFAs is $P_{ASE}$. The OSNR at the output is defined as the ratio of the optical signal power to the optical noise:
\begin{equation}
OSNR_{out}=\frac{P_{out}}{P_{ASE}}=\frac{P_{in}G}{P_{ASE}}
\label{OSNRtotal}
\end{equation}
The total ASE noise power of the cascaded EDFAs can be expressed as \cite{gumaste2003dwdm}:
\begin{equation}
P_{ASE}=F_{sys}h \nu B_{0}G
\end{equation}
where $h$ is the Planck constant, $\nu$ is the signal frequency, and $B_{0}$ is the optical bandwidth, $F_{sys}$ is cascaded equivalent NF of the system, which can be obtained as follows:
\begin{equation}
F_{sys}= {\textstyle \sum_{j=1}^{N}} \frac{F_{j}-L_{j-1}}{{\textstyle \prod_{m=1}^{j-1}} \Delta _{m} } 
\end{equation}
where $F_{j}$ is the noise figure (NF) of the j-th EDFA, $L_{j}$ is the fiber attenuations between the j-th EDFAs, $\Delta _{m}$ is product of EDFA gain and fiber link attenuation. Finally, the OSNR can be expressed as \cite{8481478}:
\begin{equation}
OSNR_{out}=\frac{P_{in}}{h\nu B_{0}} \frac{1}{F_{sys}}
\label{OSNR}
\end{equation}
According to Eq. \ref{OSNRtotal} and Eq. \ref{OSNR}, the smaller $B_{0}$ is, the higher OSNR is. $B_{0}$ is the integral bandwidth used in the calculation of the $P_{ASE}$. In the presence of an optical filter at the receiver, noise outside the filter bandwidth is effectively suppressed. Consequently, $B_{0}$ can be regarded as the bandwidth of the optical filter, and the OSNR can be improved by reducing the bandwidth of the optical filter. 

\par
\begin{figure}[h!]
\centering\includegraphics[width=0.8\columnwidth]{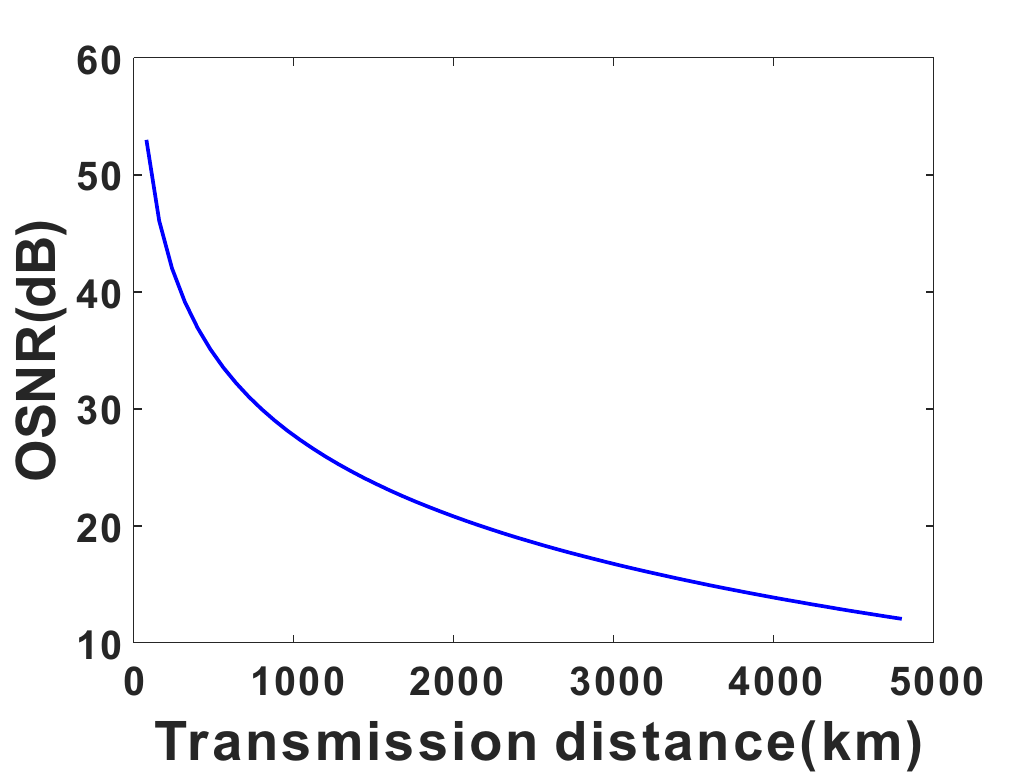}
\caption{Simulation of the OSNR of the signal under different number of EDFAs and transmission distance.}
\label{simOSNR}
\end{figure}

 Assuming that the optical wavelength is 1550 nm, $B_{0}$ is 0.1nm, $P_{in}$ is 4 dBm, every 80 km fiber to put an EDFA. The fiber attenuation coefficient is 0.2 dB/km, and the gain of each EDFA is 16dB. The NF of each EDFA in the link is the same as 5. The gain of each EDFA can fully compensate for the loss of the previous fiber segment. According to Eq. \ref{OSNR}, simulation of the OSNR of the signal under different numbers of EDFAs and transmission distance is shown in Fig. \ref{simOSNR}. In practice, due to differences in amplifiers, environmental disturbances, and link loss mutations, the OSNR will be worse than under ideal conditions. 

\begin{figure*}[h]
\centering\includegraphics[width=2\columnwidth]{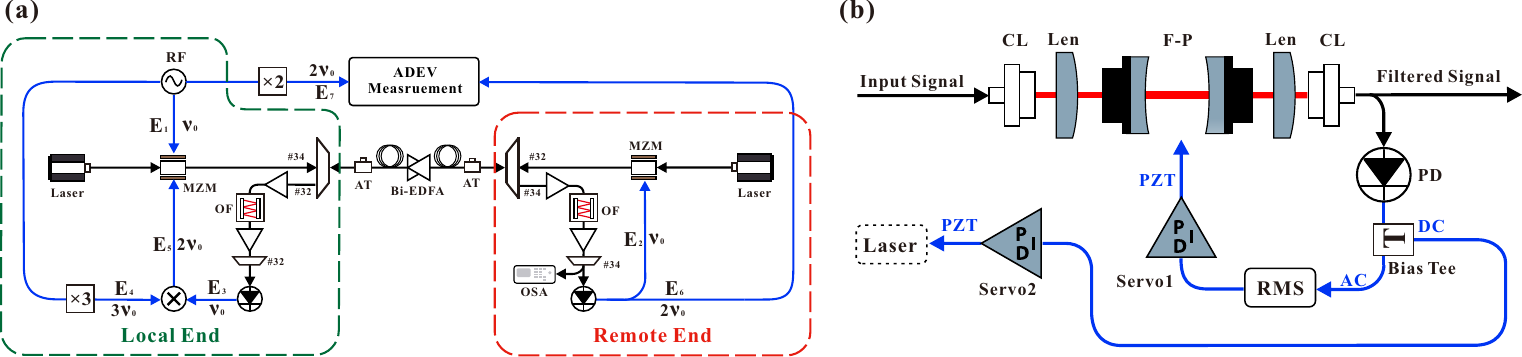}
\caption{(a) Schematic of fiber-optic radio frequency transfer based on passive phase compensation with optical filtering. MZM: Machzehnder modulator, Bi-EDFA: bidirectional erbium-doped optical fiber amplifier, AT: attenuator, OF: optical filter. PD: photoelectric detector, RF: microwave source, OSA: optical spectrum analyzer,. (b) Schematic of the F-P cavity filter system. CL: collimator, Len: focusing len, F-P: Fabry-Perot cavity, PZT: piezoelectric ceramic transducer, RMS: root-mean-square detector}
\label{structure}
\end{figure*}

\subsection{SNR and Allan Deviation}
OSNR evaluates the quality of the optical signal, but ultimately the signal must be converted to an electrical signal through PD. Therefore, the usability of the signal is determined by the SNR of the electrical signal. To calculate the electrical SNR of the amplified signal, we should add the contribution of ASE to the receiver noise. The photocurrent converted after the signal enters the PD is
\begin{equation}
I_{d}=\Re P_{signal}+{i_{b} +i_{ASE}+i_{s}+i_{r}} 
\end{equation}
where $\Re $ is the responsivity of the photodiode, $P_{signal}$ is the desired signal, $i_{b}$ and $i_{ASE}$ represent current fluctuations resulting from signal-ASE beating and ASE-ASE beating, $i_{s}$ and $i_{T}$ are current fluctuations induced by shot and thermal noises, $i_{r}$ is relative intensity noise (RIN) of laser, $R$ is the responsivity of the PD. Since these noise terms fluctuate with time, we need to find their variances. Therefore, the SNR can be expressed \cite{agrawal2005theory}:
\begin{equation}
SNR=\frac{(\Re P_{signal})^2}{\sigma _{b}^2 +\sigma _{s}^2+\sigma _{ASE}^2+\sigma _{T}^2+\sigma _{r}^2} 
\label{SNR}
\end{equation}

Eq. (\ref{OSNRtotal}) shows that a higher ASE power ratio results in a lower OSNR. Similarly, Eq. (\ref{SNR}) indicates that a higher power of ASE leads to a higher beat frequency noise including signal-ASE and ASE-ASE beating, and a reduction in SNR. Therefore, improving the OSNR will enhance the SNR of the electrical signal. Assuming only white phase noise, the Allan deviation denoting the instability can be obtained as \cite{primas1989applications,mu2021modeling,enzer2021allan}:
\begin{equation}
\sigma(\tau ) =\frac{\sqrt{3} }{2\pi \nu _{0}\tau m\sqrt{SNR}  } 
\label{allan and snr}
\end{equation}
where $m$ is the modulation depth, $\nu _{0}$ modulation frequency. As illustrated in Eq. \ref{allan and snr}, the SNR is enhanced and the Allan deviation is decreased when the modulation frequency and depth are maintained at fixed values.

\section{Experimental Setup}

Fig. \ref{structure}(a) illustrates a classic passive phase compensation system. The frequency of the microwave source output signal $\nu_{0}$ is 1.5 GHz. The total length of the fiber link is 160 km. The link loss is compensated using a bidirectional EDFA. To simulate the low OSNR condition caused by long-distance fiber transmission, we introduced two optical attenuators(AT) with 14 dB attenuation in the fiber link. The output power of both lasers is set to 0 dBm, and after transmission over the link, the light received at the local and remote ends is only -53 dBm. The gain of the first-stage unidirectional EDFA is turned up so that the signal accumulates enough optical noise. An optical filter is used to filter out the optical noise. The optical filters used in this experiment are DWDM, Waveshaper and the proposed F-P cavity filter system. The bandwidths of these filters are 100 GHz, 10 GHz, and 6 MHz, respectively. The second stage unidirectional EDFA is set to low gain, which serves to make the input optical power of the PD -1.5 dBm while avoiding introducing too much optical noise. The optical noise introduced by the second stage amplification is filtered using a DWDM single-channel filter. Fig. \ref{structure}(b) illustrates the proposed F-P cavity filtering system. The system incorporates two servo controllers, which are responsible for locking the laser frequency to the center of the F-P cavity transmission peak and stabilizing the FSR of the F-P cavity to 1.5 GHz, respectively. Both The servo controls of the laser frequency and the FSR of the F-P cavity is through the piezoelectric ceramic transducer (PZT) inside them.
\begin{figure}[t!]
\centering\includegraphics[width=0.9\columnwidth]{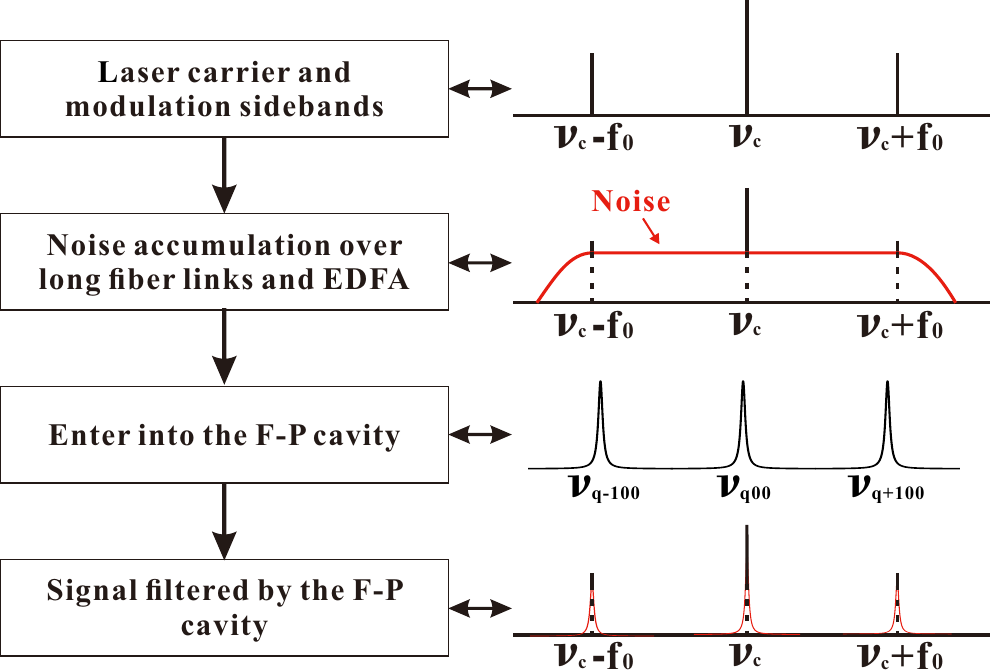}
\caption{The process of filtering and noise reduction of the RF intensity modulated signal. $\nu_{c}$: laser carrier; $\nu_{c}$-$f_{0}$, $\nu_{c}$+$f_{0}$: intensity modulation sideband.}
\label{principle}
\end{figure}
\par Fig. \ref{principle} illustrates the filtering principle based on the F-P cavity. After MZM, the sinal enters the link for transfer. Due to the attenuation of the link and the accumulation of EDFA noise, the OSNR of both the carrier and the sidebands will gradually decrease. The F-P cavity has comb filtering characteristics, so it can be used to filter the carrier and the sidebands simultaneously. In theory, the narrower the transmission peak of F-P cavity, the more noise is filtered out. 
\begin{figure*}[h!]
\centering\includegraphics[width=2\columnwidth]{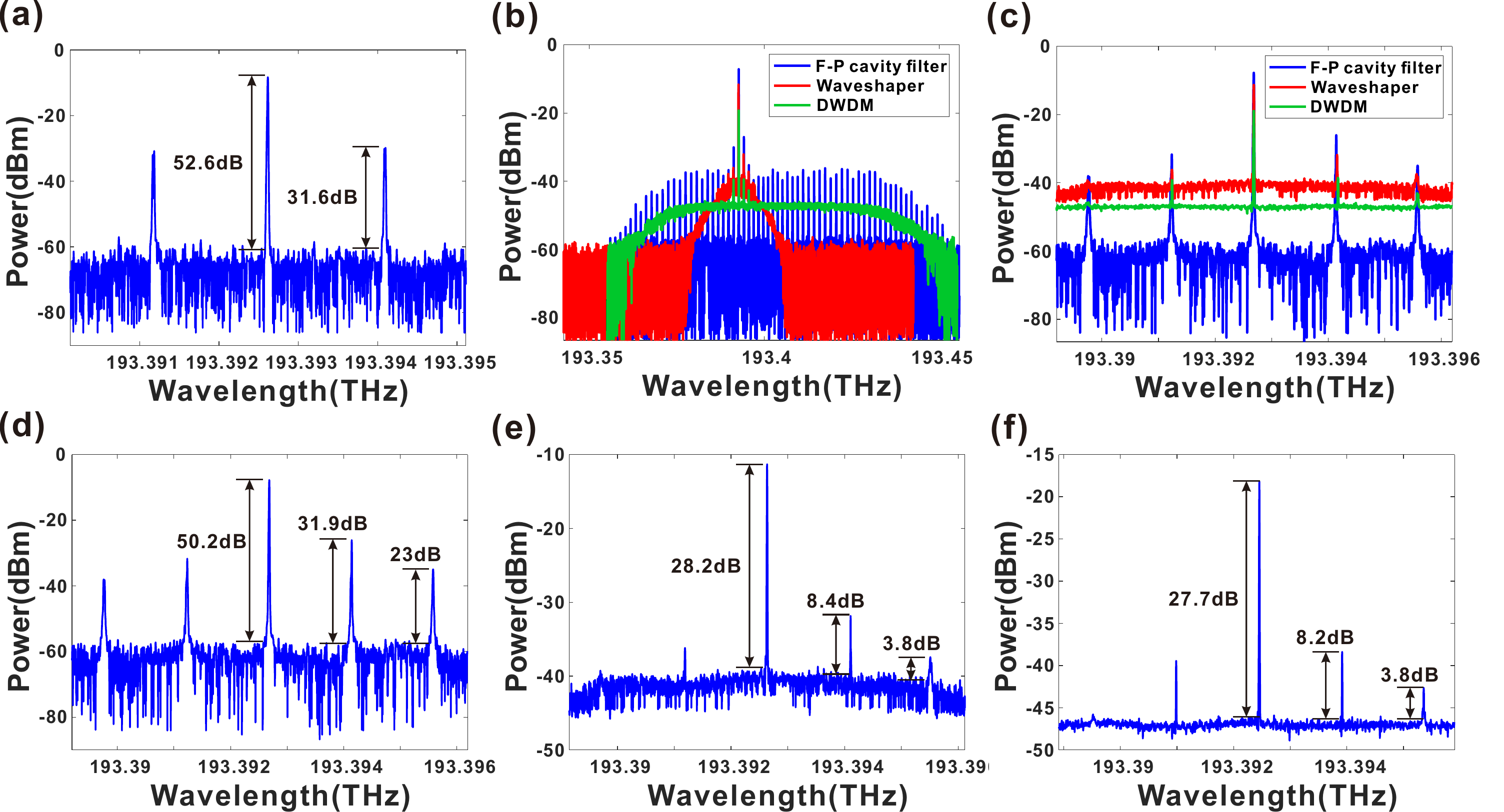}
\caption{(a) The optical spectrum of the signal to be transfer in local end, span = 5 GHz. (b) The optical spectrum comparison of the signal received at the remote end when utilising different optical filters, span = 7 GHz, (c) span = 100 GHz. (d) The optical spectrum of signal received at the remote end using F-P cavity filter, (e) using Waveshaper, (f) using DWDM, span = 7 GHz. BOSA high-resolution spectrum analyzer with a frequency resolution of 100 MHz. All instruments used in the test are BOSA high-resolution spectrum analyzer with a resolution of 10 MHz, and the optical power of the signal to be tested is -1.5 dBm.}
\label{optical spectrum}
\end{figure*}
\section{Results}
Fig. \ref{optical spectrum}(a) illustrates the optical spectrum of the intensity modulated signal at local end. Fig. \ref{optical spectrum}(b) and (c) shows the optical spectrum of the signal received at the remote end when utilising different optical filters. The spans are 100 GHz and 10 GHz, respectively. From the Fig. \ref{optical spectrum}(b), the signal is not pure but carries ASE noise. The spectrum measurements are based on the same input optical power. In this case, the larger the carried ASE noise, the smaller the signal proportion. The power of desired signal from high to low is F-P cavity, Waveshaper and DWDM. The filters with a small bandwidth can filter out more noise, resulting in a higher proportion of signal. The F-P cavity has a $\Gamma _{FWHM}$ of 6 MHz and a high rejection ratio, so signal filtered by the F-P cavity has the highest power and the lowest floor noise. In addition, it can filter out most of the noise of the first-stage EDFA, so that the second-stage EDFA can fully amplify the signal component. In Fig. \ref{optical spectrum}(b), the F-P cavity filtered signal optical spectrum appears a comb shape, which is a result of the ASE noise filtering of the EDFA by the F-P cavity. Fig. \ref{optical spectrum} (d), (e) and (f) shows the signal spectrum with F-P cavity, Waveshaper and DWDM, respectively. The span all are 7 GHz. The carrier OSNR of F-P cavity filtered signal is 2.4 dB lower than the original signal (Fig. \ref{optical spectrum}(a)) and 12 dB higher than DWDM and Waveshaper. The OSNR of 1.5 GHz sideband is close to the original signal and 23.5 dB higher than DWDM and Waveshaper. The OSNR of 3 GHz sideband is 19.2 dB higher than DWDM and Waveshaper. The floor noise does not increase significantly compared to the original signal. The filtered carrier and sideband are narrow at the top and wide at the bottom of the spectrum, and the total width is wider than the original signal. This is because the $\Gamma _{FWHM}$ of the F-P cavity is 6 MHz, which is larger than the linewidth of the carrier and sidebands, so there is still a certain amount of noise remaining. The portion not submerged by noise is narrow, the portion submerged by noise but filtered by F-P cavity is wide, and the F-P cavity transmission peak itself is also narrow at the top and wide at the bottom. Choosing an F-P cavity with a narrower transmission peak can reduce the residual noise, but it can also make locking more difficult.
\par
\begin{figure*}[ht!]
\centering\includegraphics[width=1.7\columnwidth]{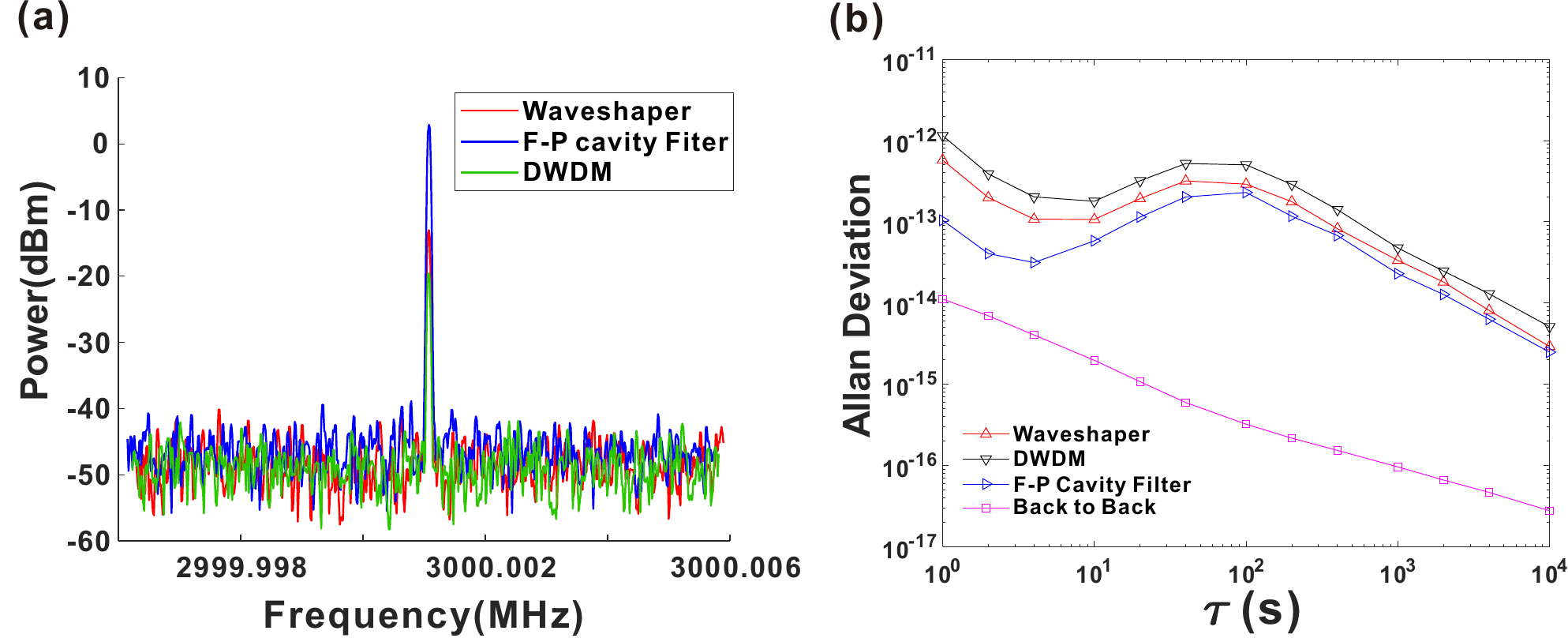}
\caption{(a) The electrical spectrum of $E_{6}$ with different filters. (b) The ADEV of the passive phase compensation system with different filters.} 
\label{snr and allan}
\end{figure*}

 Fig. \ref{snr and allan}(a) shows the electrical spectrum of $E_{6}$. The span of spectrum analyzer is 10 kHz and resolution  is 100 Hz. The input optical power to the PD is the same using different filters. The SNR of the F-P cavity, Waveshaper, and DWDM filtered signal is 39.1 dB, 28 dB, and 24.9 dB, respectively. Compared with DWDM and Waveshaper, the SNR of the F-P cavity-filtered signal is improved by 14.2 and 11.1 dB, respectively. Fig. \ref{snr and allan}(b) shows the ADEV of passive phase compensation system with different optical filters. The ADEV ($\tau =1s$) of DWDM, Waveshaper, and F-P cavity filtering is 1.15E-12, 5.74E-13, and 1.04E-13, respectively. The ADEV ($\tau =1s$) of the F-P cavity filtering improves by half an order of magnitude compared to Waveshaper and by one order of magnitude compared to DWDM. In terms of long-term stability, the F-P cavity filtering does not significantly improve compared with the other two filters. The ADEV has an obvious bulge around $\tau =100s$, which indicates that the system still has uncompensated phase fluctuation. These phase fluctuations arise from the crosstalk between signals with frequencies $\nu _{0}$ and $2\nu _{0}$. The crosstalks are mainly due to inadequate rejection ratios of filters and mixers, spatial interference due to poor electromagnetic shielding, etc. 

\section{Double-Servo Control}
The F-P cavity's comb filtering characteristics allow for the effective removal of noise surrounding the carrier and sidebands. It is possible that the carrier and sidebands will not reach the maximum transmittance of the F-P cavity simultaneously during the filtering process. This is due to the wavelength drift of the laser and the longitudinal mode drift of the F-P cavity. As a result, the filtered signal may become unstable. In Fig. \ref{structure}(b) a dual servo locking scheme is proposed to improve filter stability. 
\par First, servo 1 is run to lock the signal wavelength at the maximum transmittance of the transmission peak of the F-P cavity. Then, servo 2 is run to control the FSR of the F-P cavity to be consistent with the modulation frequency $\nu_{0}$. The DC output of the bias tee represents the intensity after the signal light passes through the F-P cavity. In this experiment, the DC value mainly represents the transmittance of the light carrier. We set the maximum DC voltage to the locking point of servo 1. The input to servo 2 is the voltage of the AC components of the bias tee after passing through the RMS detector. After running servo 1, the carrier frequency is in the center of the transmission peak. When the FSR of the F-P cavity matches the modulation frequency, the sideband transmittance is the highest. The RMS detector converts the power of the AC signal to the voltage, and servo 2 locked the voltage to the maximum. To ensure that the carrier frequency is at the center of the transmission peak when adjusting the FSR of the F-P cavity, the feedback speed of servo 1 is set much higher than that of servo 2.
\begin{figure}[tb]
\centering\includegraphics[width=1\columnwidth]{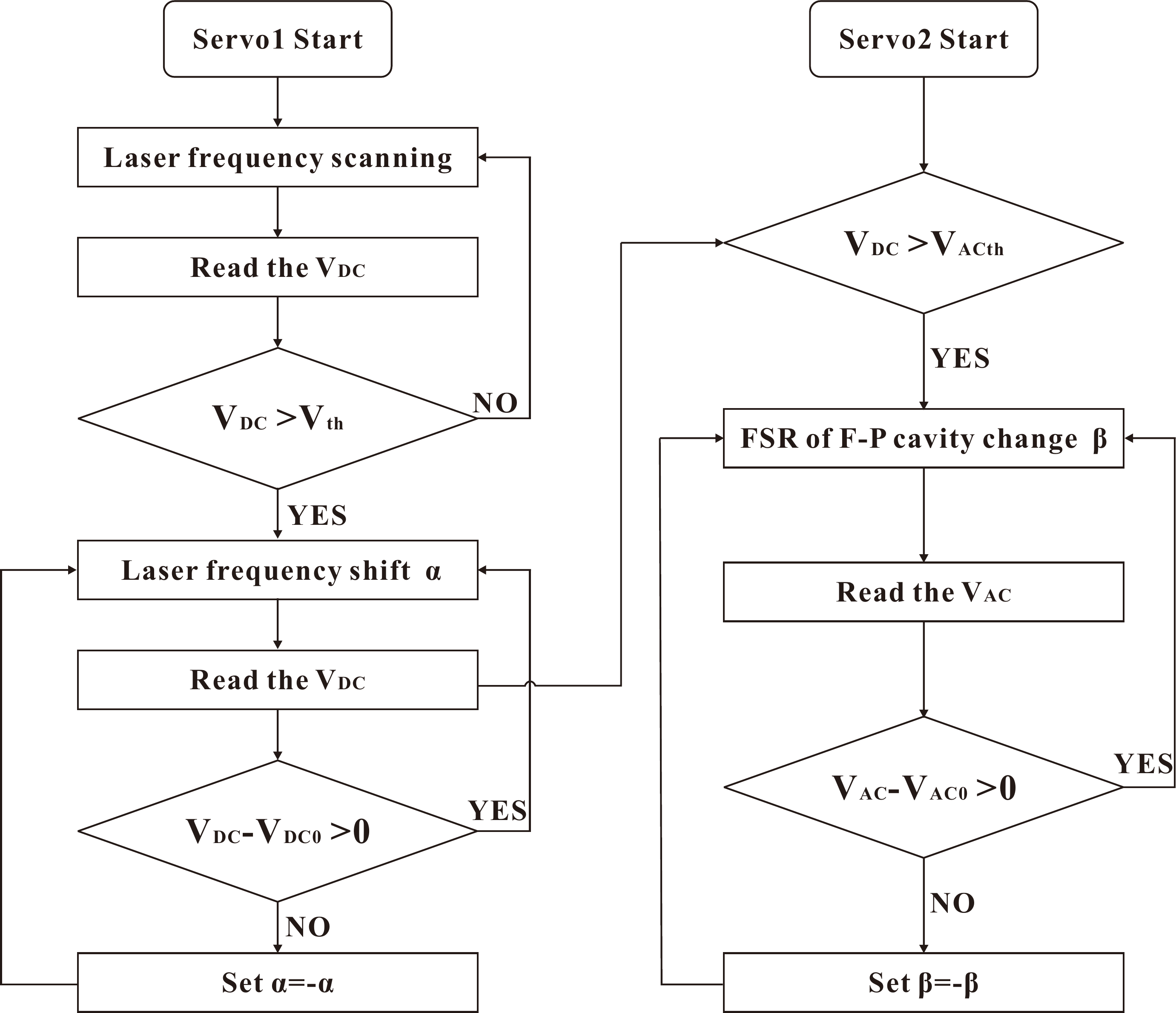}
\caption{Flow chart of double-servo control algorithm}
\label{flow chart}
\end{figure}
\begin{figure}[ht]
\centering\includegraphics[width=1\columnwidth]{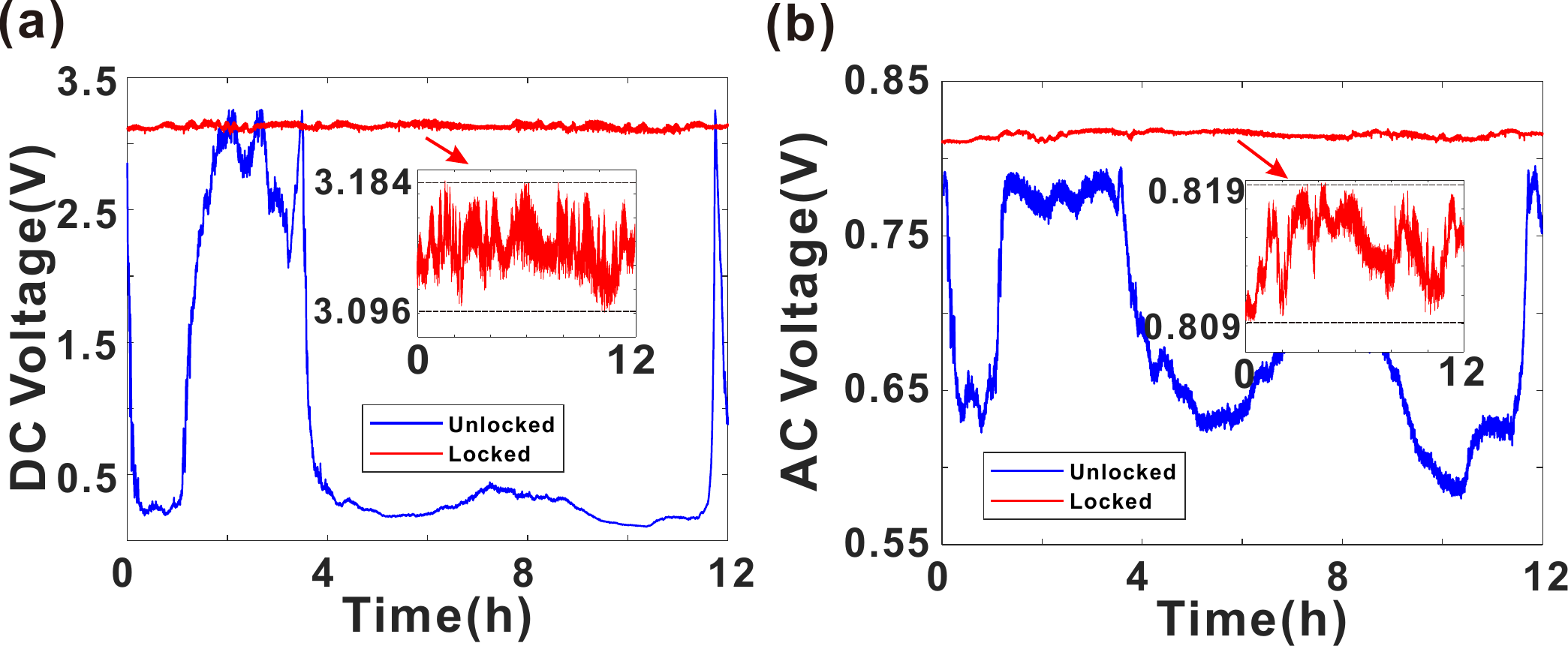}
\caption{Voltage comparison before and after locking in 12 hours, (a) DC component, (b) AC component through the RMS detector.}
\label{locking}
\end{figure}

\par The locking process is illustrated in Fig. \ref{flow chart}. In this experiment, both servo 1 and servo 2 is digital feedback. The feedback devices are the PZT in both the laser and F-P cavity. An algorithm was developed to automatically lock the maximum transmittance point. The algorithm calculates the trend of voltage variation and adjusts the voltage to reach the maximum point through feedback. Initially, servo 1 controls the laser frequency scanning and read the DC voltage. We set the current DC voltage to V$_{DC}$. The locking begins when the V$_{DC}$ reaches the threshold voltage V$_{th}$. For each feedback, read the current V$_{DC}$ and set the previous V$_{DC}$ as V$_{DC0}$. If the V$_{DC}$ - V$_{DC0}$ > 0, it indicates that the feedback direction is correct and the carrier transmittance is increasing. The next feedback will continue in this direction. If V$_{DC}$ - V$_{DC0}$ < 0, it means that the feedback direction is incorrect and the next feedback direction should be reversed. Whether servo 2 operates depends on whether the V$_{DC}$ output of servo 1 reaches V$_{ACth}$. If it does, it indicates that the carrier is close to maximum transmittance. Servo 2 functions similarly to servo 1, with the RMS detector output set to V$_{AC}$, the voltage difference between two feedbacks calculated, and the value locked near zero. 

\par Fig. \ref{locking} (a) and (b) show the locking results for servo 1 and servo 2, respectively. Both DC and AC voltages change significantly during free operation of the laser and F-P cavity. As illustrated by the blue line in Fig. \ref{locking}(a), the laser wavelength drifted to the maximum transmittance of the F-P cavity, where the transmitted light intensity was at its highest and the voltage was at its peak. When the wavelength deviates from the F-P cavity transmission peak range, the voltage rapidly decreases. Servo 1 locks the DC voltage to the maximum, as shown by the red line in Fig. \ref{locking}(a). The $V_{pp}$ is reduced from 3 V to 0.09 V in 12 hours, indicating a light transmittance greater than 98.8\% (see Appendix B for Eq. \label{transmission}). Fig. \ref{locking} (b) shows the result of servo 2, with $V_{pp}$ decreasing from 0.245 V to 0.01 V after locking, indicating a light transmittance greater than 97.9\%. Both servos 1 and 2 achieve long-term locking with excellent stability.
\section{PERFORMANCE LIMITS}
Once the F-P cavity has been selected, the performance of the locking system will determine the SNR and ADEV of the signal. During the locking process, the introduction of power fluctuation is inevitable. This power fluctuation comes from the fluctuation of the signal transmittance, while the power of the ASE is constant. The poorer the performance of the locking system, the greater the power fluctuation will be. When measuring ADEV, phase $\theta$ is usually calculated by capturing the DC voltage after signal mixing:
\begin{equation}
\theta =\arccos(\frac{V-\frac{V_{max}+V_{min} }{2} }{\frac{V_{max}-V_{min} }{2}})
\label{voltage to phase}
\end{equation}
where V is the current voltage collected by the voltmeter. By phase shifting, the maximum voltage $V_{max}$ and the minimum voltage $V_{min}$ of voltage can be measured. The allan variance can be expressed as:
\begin{equation}
\sigma^{2}(\tau)=\frac{1}{2 \tau^{2}(N-2 n)} \sum_{k=1}^{N-2 n}\left(\theta_{k+2 n}-2 \theta_{k+n}+\theta_{k}\right)^{2}
\label{phase to adev}
\end{equation}
where n is the integer multiplier of the sampling period, N is the number of data points. From Eq. \ref{voltage to phase} and \ref{phase to adev}, the power fluctuation will directly affect the voltage accuracy and thus the calculation result of ADEV. 
\begin{figure}[ht]
\centering\includegraphics[width=1\columnwidth]{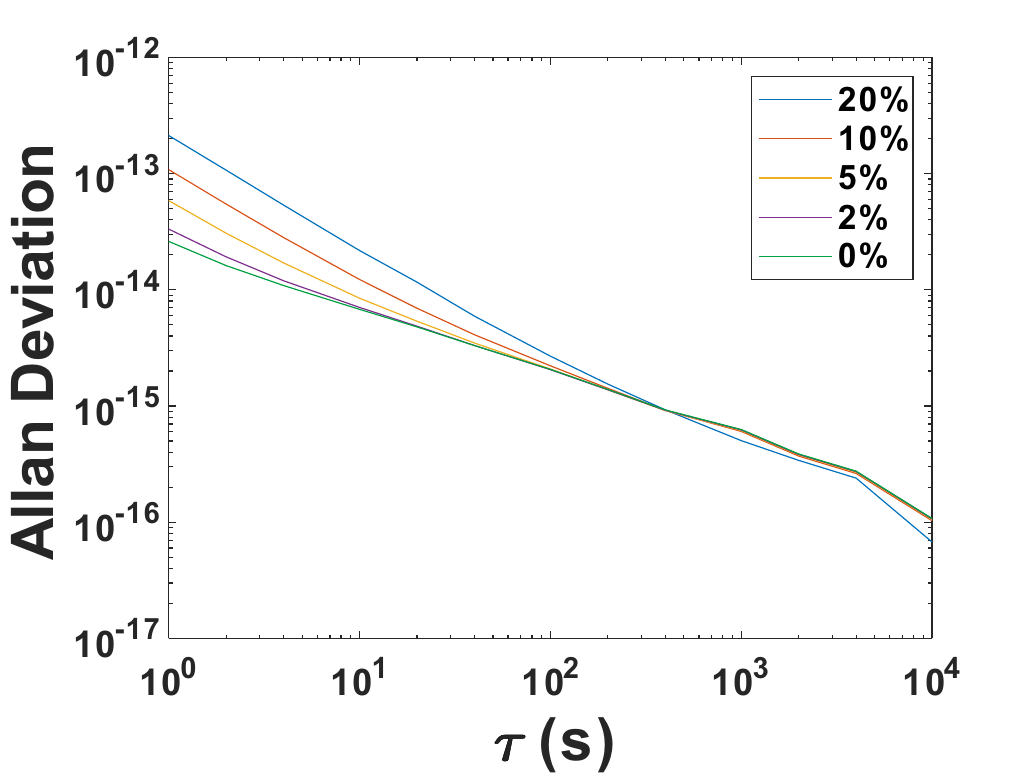}
\caption{The simulation of signal power fluctuation versus ADEV.}
\label{poweradev}
\end{figure}
\par The simulation of power fluctuation versus ADEV is shown in Fig. \ref{poweradev}. The simulation is based on real data from back-to-back tests of the passive compensation system shown in Fig. \ref{structure}(a). The back-to-back tests did not introduce optical noise, so voltage fluctuations can be approximated as power fluctuations in the signal. Assuming that the signal optical power fluctuation is $\alpha P_{signal} $, the photocurrent fluctuation is $\Re\alpha P_{signal} $ after entering the PD. $\alpha$ is fluctuation range. This fluctuation will be superimposed directly on the coefficients of $E_{6}$, which will be superimposed on the DC voltage after $E_{6}$ is mixed with $E_{7}$. Therefore, we can simulate signal power fluctuation by multiplying the real voltage data by a random array of numbers in the interval $[1-\alpha, 1]$. The sampling interval of the voltmeter is 1 s. Fig. \ref{poweradev} illustrates the variation of ADEV at different $\alpha$. As the fluctuation range increases, the short-term stability of the system deteriorates gradually. When the fluctuation range increases to 20 \%, the ADEV ($\tau = 1 s$) deteriorates by an order of magnitude. In addition, the increase in fluctuation range has little effect on the long-term stability.
\par According to the filtering principle of the F-P cavity, the narrower $\Gamma _{FWHM}$ of the filter, the less ASE remains. When the length of the F-P cavity is fixed, the fineness is the determining factor in $\Gamma _{FWHM}$, as illustrated by Eq. \ref{FWHM}. When the fineness becomes $1/N$ times, the $\Gamma _{FWHM}$ becomes N times and the residual ASE power becomes N times. As a result, the ASE-signal beating power becomes $N^2$ times and the ASE-ASE beating power becomes $N^4$ times. In FP cavity filtering, the residual ASE power is much lower than the signal power, so the power of the ASE self beating can be approximately neglected. Assuming that the shot and thermal noise of the PD and the RIN of the laser are small enough to be neglected, the SNR can be approximated to become $1/N^2$ times according to Eq. \ref{SNR}. Subsequently, the relationship between F-P cavity fineness and ADEV can be determined according to Eq. \ref{snr and allan}:
\begin{equation}
\sigma(\tau)\propto \frac{1}{F}
\end{equation}
where F is fineness of F-P cavity. Under the premise of no locking error, the F-P cavity fineness is inversely proportional to ADEV. Therefore, the ADEV can be reduced by improving the F-P cavity fineness.

\section{CONCLUSION}
This paper proposes an adaptive OSNR recovery scheme based on an F-P cavity for low OSNR. The scheme accurately filters out the noise around the laser carrier and sidebands. In the experiment of RF transfer over optical fiber, it is demonstrated that, in comparison to DWDM and Waveshaper, the F-P cavity filtering can enhance the OSNR of the carrier and sidebands by approximately 12 dB and 23.5 dB, respectively. Furthermore, the OSNR is observed to be nearly equivalent to that of the signal after modulation. The ADEV of the F-P cavity filter exhibiting an order of magnitude and a half-order of magnitude improvement over DWDM and Waveshaper at average time of 1 s, respectively. This scheme demonstrates effective noise filtering and an improvement in SNR in both electricity and light. This scheme can be combined with other schemes to optimize the signal and extend the transmission distance. In the future, this scheme can be integrated into a chip to improve compactness and ease of use.
\begin{appendices}
\section*{Appendix A: Principle of the Passive Compensation Scheme}
In order to eliminate the fluctuation of the fiber link, a passive compensation method based on round-trip phase correction is used in the RF dissemination system. At the local end in Fig. \ref{structure}(a), the RF source signal $E_{1}$ can be expressed as $E_{1} = cos(2\pi\nu_{0}t+\varphi _{0}) $, where $\nu_{0}$ and $\varphi _{0}$ are the standard frequency and original phase of the source signal, respectively. It is also the reference signal compared to the signal received at the remote end to measure the transmission stability of the RF dissemination system. The signal $E_{1}$ is modulated onto the optical carrier and transmitted in the fiber link. Due to the influence of the external environment, the phase fluctuation $\varphi _{p}$ caused by the fiber link will be added to the transmitted signal. Thus, the probe signal is received by a PD to get the uncompensated signal $E_{2}$ which can be written as:
\begin{equation}
E_{2} = cos(2\pi\nu_{0}t+\varphi _{0}+\varphi _{p})
\end{equation}
However, $\varphi _{p}$ changes with random temperature and mechanical vibration and $E_{2}$ cannot be used as an ultra-stable RF signal. According to the principle of round-trip correction, it is assumed that the phase fluctuation introduced by forward and backward transmission is equal \cite{foreman2007remote,primas1988fiber}. That is to say, the phase fluctuation caused by the backward transmission is also $\varphi _{p}$. The round-trip RF signal $E_{3}$ detected at the local end can be expressed as:
\begin{equation}
E_{3} = cos(2\pi\nu_{0}t+\varphi _{0}+2\varphi _{p})
\end{equation}
After round trip transmission, the triple frequency signals of $E_{1}$ is mixed with the returned signal $E_{3}$. The mixed signal is bandpass filtered to obtain a phase-conjugated signal $E_{5}$.
\begin{equation}
E_{5} = cos(2\pi\nu_{0}t+2\varphi _{0}-2\varphi _{p})
\end{equation}
The $E_{1}$ and $E_{5}$ are simultaneously modulated by the MZM and is transmitted to the remote end again. The compensated signal $E_{6}$ can be expressed as
\begin{equation}
E_{6} = cos(2\pi\nu_{0}t+2\varphi _{0})
\end{equation}
$E_{6}$ will be compared with the $E_{7}$ to measure the transmission stability of the system.
\section*{Appendix B: Simulation of F-P cavity transmission spectrum}
\begin{figure}[tb!]
\centering\includegraphics[width=0.9\columnwidth]{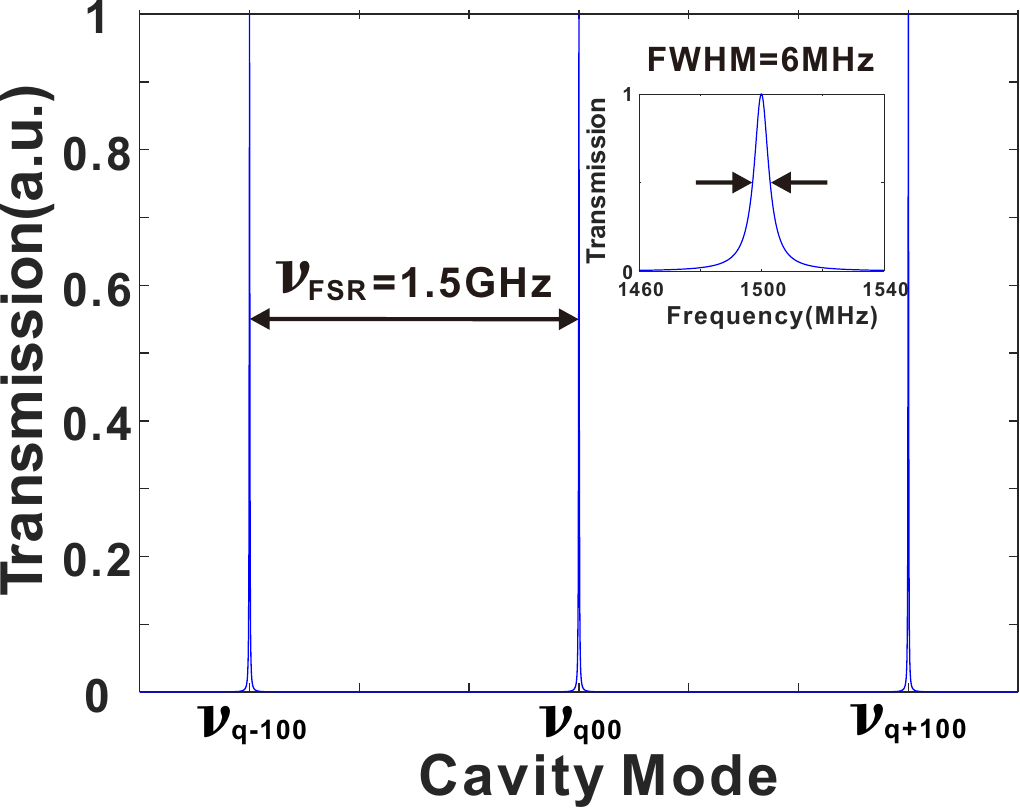}
\caption{Simulation of mode spectrum of the used Fabry-Perot cavity. ${\nu }_{FSR}$: free spectral range; $\nu_{q-100}$ $\nu_{q00}$ $\nu_{q+100}$: three consecutive TEM$_{00}$ modes.}
\label{simfp}
\end{figure}
The Fabry-Perot (F-P) cavity is a comb filter, with a narrow filtering bandwidth achieved through careful design of the cavity structure. This makes it well-suited for noise suppression. When the wave fronts of a Gaussian beam perfectly match the mirror surfaces and the incoming beam is aligned to the optical axis of the F-P resonator, its transmission spectrum is exclusively composed of TEM$_{00}$ modes that vary depending on the parameter q. The free spectral range (FSR) of the resonator is the distance between two consecutive TEM$_{00}$ modes and is given by

\begin{equation}
{\nu }_{FSR}=\frac{c}{2L}   
\end{equation}
where c is the speed of light and L is the cavity length. This equation applies to all two-mirror linear resonators. A resonator's transmission intensity, $I_{t}$, as a function of frequency detuning from mode q, $\Delta \nu _{q} =\nu -\nu _{q}$, is given by the Airy formula \cite{stephan1996airy}:
\begin{equation}
I_{t} (\Delta \nu _{q} )=\frac{t_{1}t_{2} }{(1-\sqrt{r_{1}r_{2}} )^{2} } \frac{I_{0}}{1+\frac{4\sqrt{r_{1}r_{2}}}{(1-\sqrt{r_{1}r_{2}} )^{2}}\sin ^{2} (\frac{\pi \Delta \nu _{q}}{\nu _{FSR}})  } 
\label{transmission}
\end{equation}
where $I_{0}$ is the intensity of the light incident on the instrument, $t_{1}$, $t_{2}$, $r_{1}$ and $r_{2}$ represent the transmission and reflection of the input and output mirror respectively. 
\par The performance of the F-P cavity is highly dependent on the reflectivity of the mirror surface. A low reflector will produce a broader transmission peak, while a high reflector will produce a narrower transmission peak. In addition to the free spectral range, there are two other important quantities for F-P cavity: fineness and mode width. The fineness $F$ with the same reflection coefficient $r$ of mirrors is
\begin{equation}
F=\frac{\pi \sqrt{r} }{1-r} 
\end{equation}
The mode width of the F-P cavity is related to fineness $F$ and FSR by
\begin{equation}
\Gamma _{FWHM}=\frac{\nu _{FSR} }{F}
\label{FWHM}
\end{equation}

\par According to Eq. (\ref{transmission}), the relationship between frequency and transmission of F-P cavity is simulated, and the results are shown in Fig. \ref{simfp}. The F-P cavity used in this paper has an FSR of 1.5 GHz and a fineness of 250. The upper right plot of Fig. \ref{simfp} shows in detail one of the TEM$_{00}$ modes, which has a $\Gamma_{FWHM}$ of 6 MHz according to Eq. (\ref{FWHM}).
\end{appendices}




\bibliography{main}

\begin{thebibliography}{10}
\newcommand{\enquote}[1]{``#1''}

\bibitem{yu2010ultra}
J.~Yu and X.~Zhou, {\protect\JournalTitle{IEEE Communications Magazine}} \textbf{48}, S56 (2010).

\bibitem{tseng2013survey}
W.-H. Tseng and S.-Y. Lin, {\protect\JournalTitle{NCSLI Measure}} \textbf{8}, 70 (2013).

\bibitem{grotti2018geodesy}
J.~Grotti, S.~Koller, S.~Vogt, \emph{et~al.}, {\protect\JournalTitle{Nature Physics}} \textbf{14}, 437 (2018).

\bibitem{essiambre2012capacity}
R.-J. Essiambre and R.~W. Tkach, {\protect\JournalTitle{Proceedings of the IEEE}} \textbf{100}, 1035 (2012).

\bibitem{predehl2012920}
K.~Predehl, G.~Grosche, S.~Raupach, \emph{et~al.}, {\protect\JournalTitle{Science}} \textbf{336}, 441 (2012).

\bibitem{droste2013optical}
S.~Droste, F.~Ozimek, T.~Udem, \emph{et~al.}, {\protect\JournalTitle{Physical review letters}} \textbf{111}, 110801 (2013).

\bibitem{bobrovs2013comparative}
V.~Bobrovs, S.~Olonkins, A.~Alsevska, \emph{et~al.}, {\protect\JournalTitle{International Journal of Physical Sciences}} \textbf{8}, 1898 (2013).

\bibitem{Terra:10}
O.~Terra, G.~Grosche, and H.~Schnatz, {\protect\JournalTitle{Opt. Express}} \textbf{18}, 16102 (2010).

\bibitem{islam2002raman}
M.~N. Islam, {\protect\JournalTitle{IEEE Journal of selected topics in Quantum Electronics}} \textbf{8}, 548 (2002).

\bibitem{okoumassoun2023performance}
T.~P. Okoumassoun, A.~Antwiwaa, and N.~K. Gerrar, {\protect\JournalTitle{Journal of Communications}} \textbf{18} (2023).

\bibitem{liu2020optical}
Z.~Liu and R.~Slav{\'\i}k, {\protect\JournalTitle{Journal of Lightwave Technology}} \textbf{38}, 43 (2020).

\bibitem{pedersen1991design}
B.~Pedersen, A.~Bjarklev, J.~H. Povlsen, \emph{et~al.}, {\protect\JournalTitle{Journal of lightwave technology}} \textbf{9}, 1105 (1991).

\bibitem{demir2007nonlinear}
A.~Demir, {\protect\JournalTitle{Journal of lightwave technology}} \textbf{25}, 2002 (2007).

\bibitem{charles1966some}
F.~Charles and W.~C. Lindsey, {\protect\JournalTitle{Proceedings of the IEEE}} \textbf{54}, 1152 (1966).

\bibitem{liu2020ultrastable}
C.~Liu, J.~Shang, Z.~Zhao, \emph{et~al.}, {\protect\JournalTitle{IEEE Photonics Journal}} \textbf{13}, 1 (2020).

\bibitem{chiou1996effect}
Y.~Chiou and L.~Wang, {\protect\JournalTitle{Journal of lightwave technology}} \textbf{14}, 2126 (1996).

\bibitem{ascheid1982cycle}
G.~Ascheid and H.~Meyr, {\protect\JournalTitle{IEEE transactions on communications}} \textbf{30}, 2228 (1982).

\bibitem{salwik2021optimization}
K.~Salwik, {\L}.~{\'S}liwczy{\'n}ski, P.~Krehlik, and J.~Ko{\l}odziej, {\protect\JournalTitle{Optical Fiber Technology}} \textbf{62}, 102465 (2021).

\bibitem{sliwczynski2012bidirectional}
{\L}.~Sliwczynski and J.~Kolodziej, {\protect\JournalTitle{IEEE Transactions on Instrumentation and Measurement}} \textbf{62}, 253 (2012).

\bibitem{yin2018reference}
G.~Yin, S.~Cui, C.~Ke, and D.~Liu, {\protect\JournalTitle{IEEE Photonics Journal}} \textbf{10}, 1 (2018).

\bibitem{weiss2003performance}
A.~J. Weiss, {\protect\JournalTitle{IEEE Photonics Technology Letters}} \textbf{15}, 1225 (2003).

\bibitem{aloisio2012performance}
A.~Aloisio, F.~Ameli, A.~D'Amico, \emph{et~al.}, {\protect\JournalTitle{IEEE transactions on nuclear science}} \textbf{59}, 251 (2012).

\bibitem{gavioli2010investigation}
G.~Gavioli, E.~Torrengo, G.~Bosco, \emph{et~al.}, \enquote{Investigation of the impact of ultra-narrow carrier spacing on the transmission of a 10-carrier 1tb/s superchannel,} in \emph{2010 Conference on Optical Fiber Communication (OFC/NFOEC), collocated National Fiber Optic Engineers Conference,}  (IEEE, 2010), pp. 1--3.

\bibitem{ozolicnvs2009realization}
O.~Ozoli{\c{n}}{\v{s}} and {\c{G}}.~Ivanovs, {\protect\JournalTitle{Elektronika ir Elektrotechnika}} \textbf{92}, 41 (2009).

\bibitem{iodice2000silicon}
M.~Iodice, G.~Cocorullo, F.~Della~Corte, and I.~Rendina, {\protect\JournalTitle{Optics communications}} \textbf{183}, 415 (2000).

\bibitem{kim2016rapidly}
H.-J. Kim, D.~E. Leaird, and A.~M. Weiner, {\protect\JournalTitle{IEEE Transactions on Microwave Theory and Techniques}} \textbf{64}, 3351 (2016).

\bibitem{Yu:24}
Y.~Yu, K.~Tian, W.~Meng, \emph{et~al.}, {\protect\JournalTitle{Opt. Express}} \textbf{32}, 7574 (2024).

\bibitem{gumaste2003dwdm}
A.~Gumaste and T.~Antony, \emph{DWDM network designs and engineering solutions} (Cisco press, 2003).

\bibitem{8481478}
Y.~Xiang, M.~Tang, Q.~Wu, \emph{et~al.}, {\protect\JournalTitle{IEEE Photonics Journal}} \textbf{10}, 1 (2018).

\bibitem{agrawal2005theory}
G.~P. Agrawal, {\protect\JournalTitle{Raman Amplification in Fiber Optical Communication Systems}} pp. 33--102 (2005).

\bibitem{primas1989applications}
L.~E. Primas, R.~T. Logan, and G.~F. Lutes, \enquote{Applications of ultra-stable fiber optic distribution systems,} in \emph{Proceedings of the 43rd Annual Symposium on Frequency Control,}  (IEEE, 1989), pp. 202--211.

\bibitem{mu2021modeling}
K.-L. Mu, X.~Chen, Z.-K. Wang, \emph{et~al.}, {\protect\JournalTitle{Chinese Physics B}} \textbf{30}, 074208 (2021).

\bibitem{enzer2021allan}
D.~G. Enzer, D.~W. Murphy, and E.~A. Burt, {\protect\JournalTitle{IEEE Transactions on Ultrasonics, Ferroelectrics, and Frequency Control}} \textbf{68}, 2590 (2021).

\bibitem{foreman2007remote}
S.~M. Foreman, K.~W. Holman, D.~D. Hudson, \emph{et~al.}, {\protect\JournalTitle{Review of Scientific Instruments}} \textbf{78} (2007).

\bibitem{primas1988fiber}
L.~Primas, G.~Lutes, and R.~Sydnor, \enquote{Fiber optic frequency transfer link,} in \emph{Proceedings of the 42nd Annual Frequency Control Symposium, 1988.},  (IEEE, 1988), pp. 478--484.

\bibitem{stephan1996airy}
G.~St{\'e}phan, {\protect\JournalTitle{Journal of Nonlinear Optical Physics \& Materials}} \textbf{5}, 551 (1996).

\end{thebibliography}

\ifthenelse{\equal{\journalref}{aop}}{%
\section*{Author Biographies}
\begingroup
\setlength\intextsep{0pt}
\begin{minipage}[t][6.3cm][t]{1.0\textwidth} 
  \begin{wrapfigure}{L}{0.25\textwidth}
    \includegraphics[width=0.25\textwidth]{john_smith.eps}
  \end{wrapfigure}
  \noindent
  {\bfseries John Smith} received his BSc (Mathematics) in 2000 from The University of Maryland. His research interests include lasers and optics.
\end{minipage}
\begin{minipage}{1.0\textwidth}
  \begin{wrapfigure}{L}{0.25\textwidth}
    \includegraphics[width=0.25\textwidth]{alice_smith.eps}
  \end{wrapfigure}
  \noindent
  {\bfseries Alice Smith} also received her BSc (Mathematics) in 2000 from The University of Maryland. Her research interests also include lasers and optics.
\end{minipage}
\endgroup
}{}

\end{document}